\documentclass[prb,preprint]{revtex4-1} 

\usepackage{amsmath, amssymb}  
\usepackage{amsfonts} 
\usepackage[dvipdfmx]{graphicx} 
\usepackage{bm}

\begin{document}


\title{Home-run probability as a function of the coefficient of restitution of baseballs}

\author{Hiroto Kuninaka}
\email{kuninaka@edu.mie-u.ac.jp} 
\author{Ikuma Kosaka}
\author{Hiroshi Mizutani}
\affiliation{Faculty of Education, Mie University, Tsu City, Mie, Japan 514-8507}



\date{\today}

\begin{abstract}
In baseball games, the coefficient of restitution of baseballs strongly affects the 
flying distance of batted balls, which determines the home-run probability.  
In Japan, the range of the coefficient of restitution of official baseballs has changed 
frequently over the past five years, causing the  number of home runs to vary drastically. 
We analyzed data from Japanese baseball games played in 2014 to investigate the statistical 
properties of pitched balls. In addition, we used the analysis results to develop 
a baseball-batting simulator for determining the home-run probability as a function of 
the coefficient of restitution. Our simulation results are explained 
by a simple theoretical argument.
\end{abstract}

\maketitle 

\section{Introduction} 
The bounce characteristics of baseballs have a large influence in baseball games; 
thus, baseball organizations often establish rules 
concerning official balls. For example, in Major League Baseball (MLB), 
the baseballs are made by tightly winding yarn around a small core and covering it 
with two strips of white horsehide or cowhide\cite{cross}.
  
For the estimation of the bounce characteristics, 
the coefficient of restitution $e$ is widely used, which is defined as
\begin{equation}
e = \frac{V_{r}}{V_{i}},
\end{equation}
where $V_{i}$ and $V_{r}$ are the speeds of incidence and rebound, respectively, 
in a head-on collision of a ball with a plane. 
Note that the coefficient of restitution determines 
the loss of translational energy during the collision. 
The coefficient of restitution depends on the kind of 
material and the internal structure of the ball, as well as other factors 
such as the impact speed\cite{cross, adair, stronge, johnson, goldsmith}, 
impact angle\cite{louge, kuninaka_prl}, temperature of the ball\cite{drane, allen}, 
and humidity at which the balls are stored\cite{kagan}.

Various baseball organizations officially determine the range of 
the coefficient of restitution of baseballs. 
For example, the coefficient of restitution of an MLB baseball is required to be 
$0.546 \pm 0.032$\cite{kagan2}. Regarding Japanese professional baseballs, 
the Nippon Professional Baseball Organization (NPB) first introduced their 
official baseball in 2011, which was used in both Pacific and Central League. 
Table \ref{tb1} shows a chronological table indicating the average coefficient 
of restitution for baseballs used in Japanese professional baseball games 
from 2010 to 2013\cite{npb}, along with the annual number of home runs\cite{brc}. 
Clearly, the number of home runs decreased drastically in 2011 compared with 2010, 
although the difference in the average coefficient of restitution is only on the order of $10^{-2}$. 
The average coefficient of restitution increased in 2013 
because the NPB made a baseball equipment manufacturer change the specification 
of the baseballs in order to increase the level of offense in baseball games.  

\begin{table}[h!]
\centering
\caption{Coefficient of restitution of baseballs and number of home runs in Japanese professional baseball games.}
\begin{ruledtabular}
\begin{tabular}{l c c c c p{5cm}}
Year & Coefficient of Restitution (average)& Number of Home Runs\\
\hline	
2010& 0.418& 1,605 \\
2011& 0.408 &  939\\
2012& 0.408 &  881\\
2013& 0.416 & 1,311
\end{tabular}
\end{ruledtabular}
\label{tb1}
\end{table}

Generally,  the number of home runs is strongly affected not only by the coefficient of restitution of 
baseballs, but also other various factors, such as the climate, the specifications of bats, 
and the batting skills of players, and so on. 
Sawicki et al. constructed a detailed batting model incorporating several factors, 
including the air resistance, friction between the bat and ball, wind velocity, 
and bat swing\cite{sawicki}. 
Although they investigated the optimal strategy for achieving the maximal range of a 
batted ball, they did not calculate the home-run probability, 
because it may be difficult to choose proper parameters for the home run probability function. 
However, a quantitative research on the relationship between the coefficient 
of restitution of baseballs and the home-run probability is valuable for two reasons. 
First, Table. \ref{tb1} indicates that the home-run probability strongly depends on 
the coefficient of restitution of baseballs because the small amount of changes 
in the coefficient of restitution can alter the flying distances of batted balls\cite{nathan11}. 
Second, the coefficient of restitution of baseballs 
is a controllable factor that is important 
for the design of baseball equipment.  
The home run probability as a function of the coefficient of restitution 
can be a simple criterion to evaluate the characteristics of 
official baseballs.
In addition, a quantitative research on 
the relationship between the coefficient of restitution of balls and the home-run 
probability is also valuable for physics education, 
as the problem is closely related to topics covered in undergraduate physics.

In this study, we developed a batting simulator using real baseball data to 
quantitatively investigate the home-run probability as a function of 
the coefficient of restitution. This paper is structured as follows. 
In the next section, we describe the data analysis and analysis results. 
Sections 3 presents the construction of our batting simulator and the simulation 
results. In Sections 4 and 5, we discuss and summarize our results. 
Appendices A and B are devoted to the derivation of the averaged force in 
a binary collision between a ball and a bat and the algorithm for the collision, respectively.

\section{Data Analysis}
\begin{figure}[h!]
\centering
\includegraphics[width=10cm]{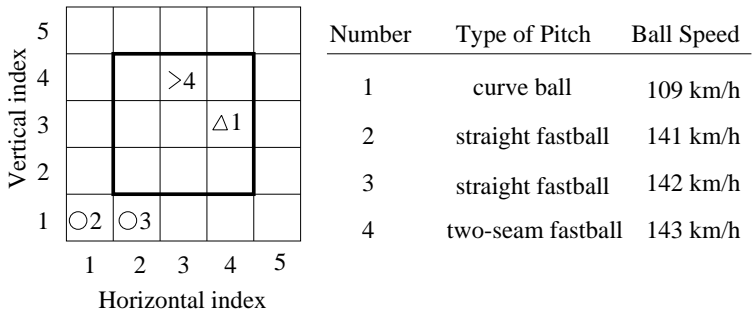}
\caption{Schematic of a part of the Sportsnavi page. }
\label{fig1}
\end{figure}

To construct our batting simulator, we first analyzed pitching data for Japanese 
professional baseball games held in 2014. 
We used data from Sportsnavi\cite{sportsnavi}, which show various data about 
the pitched balls in an official game, including the ball speed, pitch type, and 
position of a ball crossing the home plate. 
Figure \ref{fig1} shows a schematic of a part of a Sportsnavi page. 
In the data, the pitching zone is divided into 5 $\times$ 5 grids 
from the pitcher's perspective, wherein 3 $\times$ 3 grids, represented by thick lines, 
corresponds to the strike zone (see the left panel of Fig. \ref{fig1}).  
The numbers and the symbols in a grid respectively show the order and types of 
pitches, respectively, at different positions on the grid. 
Information about each pitch, including the ball speed, is presented in the table 
shown in the right side of Fig. \ref{fig1}. 
For a later discussion, we numbered the horizontal and vertical positions of 
each grid as shown in Fig. \ref{fig1}.

Using the Sportsnavi database, we manually recorded all the positions 
and ball speeds of pitches in 12 selected games held in Nagoya Dome Stadium 
in Nagoya, Japan, from August 6 2014 to September 25 2014. 
We chose games held in indoor domes because the flight of baseballs is hardly 
affected by climatic factors such as the wind strength. 
We collected and analyzed data for 1,548 pitched balls.

\begin{figure}[h!]
\centering
\includegraphics[width=7cm]{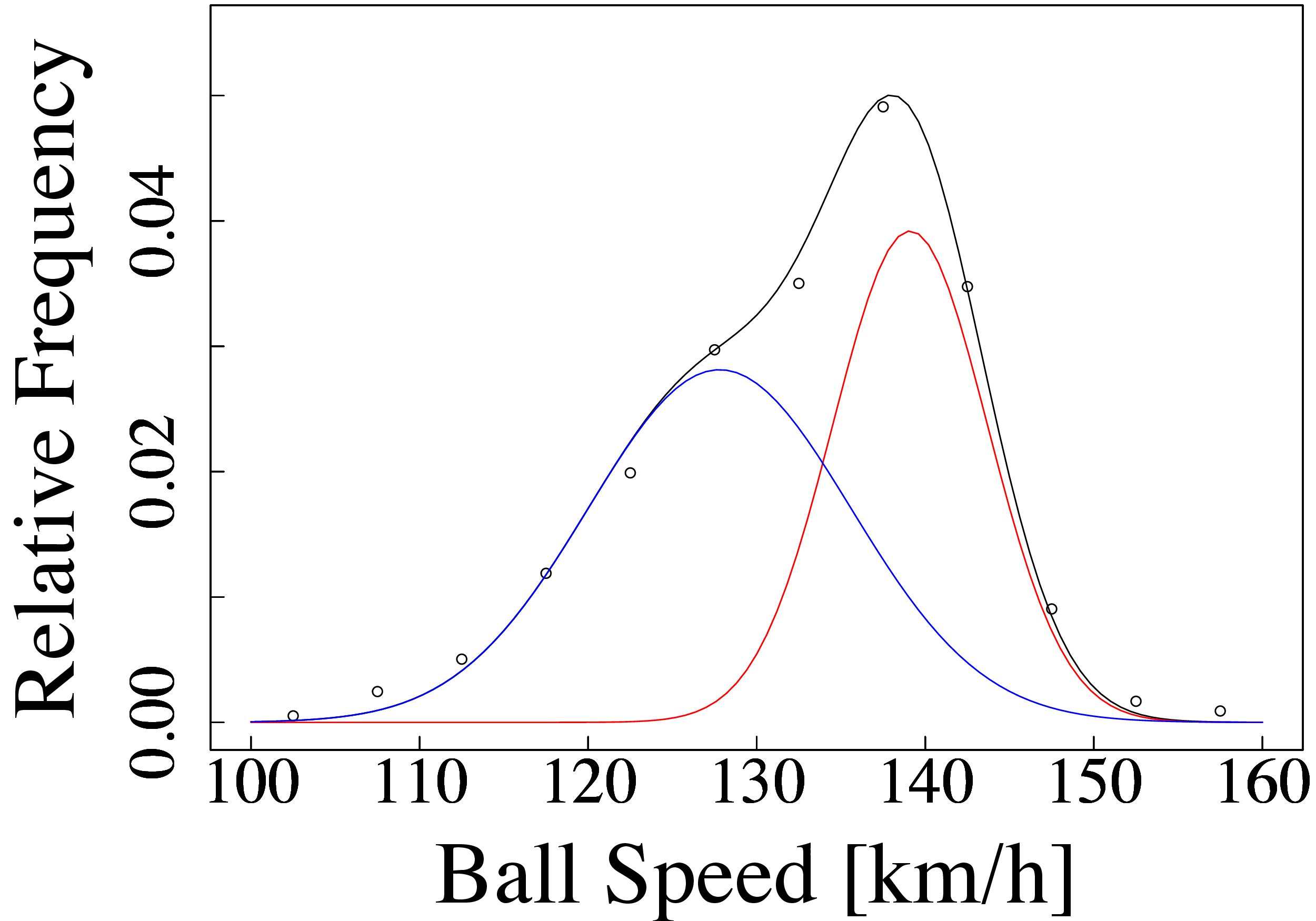}
\caption{Distribution of ball speeds. Open circles show the probabilities at each ball speed. 
Solid black curve shows Eq. (\ref{md}) with the fitting parameters shown in Table \ref{tb2}. 
Solid red and blue curves show the distributions of the straight and the breaking balls, 
respectively, weighted with $p=0.45$.} 
\label{fig2}
\end{figure}

Figure \ref{fig2} shows the distribution of the pitched-ball speed $v$, where 
the open circles indicate the calculated probabilities as a function of $v$. 
To obtain the distribution function approximating these data, 
we divided the ball-speed data into two categories: those for straight balls 
and those for breaking balls having a curve, a two-seam fastball, etc. 
For each of the categorized data points, we fit the normal distribution defined by 
\begin{equation}
f_{i}(v) = \frac{1}{\sqrt{2 \pi }\sigma_{i}} \exp \left\{ 
- \frac{(v - \mu_{i})^{2}}{\sigma_{i}^{2}} \right\} \hspace{3mm} (i = 1, 2), \label{norm}
\end{equation}
where $\mu_{i}$ and $\sigma_{i}$ are the mean and the standard deviation, respectively. 
The fitting parameters are presented in Table \ref{tb2}, where $i=1$ and $i=2$ 
correspond to the straight and breaking balls, respectively.

Considering $f_{i}(v)$ ($i=1, 2$) to be components, 
we finally obtained the mixture distribution of the pitched-ball speed $v$ as
\begin{eqnarray}\label{md}
\phi(v) = p f_{1}(v) + (1-p) f_{2}(v) \hspace{3mm} (0 < p < 1).
\end{eqnarray} 
Here, $p$ is the mixing parameter. The black solid curve shown in Fig.\ref{fig2} 
indicates Eq. (\ref{md}) with $p=0.45$, which closely approximates  
the distribution of the pitched-ball speed with the coefficient of determination equal 
to 0.9942. 
The red and the blue curves show the first and the second terms, respectively, 
in the right-hand side of Eq. (\ref{md}). Generally, the value of $p$ represents 
the probability of selecting each component of the mixture distribution. 
Thus, we consider that the pitchers chose straight and breaking balls with probabilities 
of $p=0.45$ and $0.55$, respectively.  

\begin{table}[h!]
\centering
\caption{Fitting parameters used in Eq. (\ref{norm}).}
\begin{ruledtabular}
\begin{tabular}{l c c c c p{5cm}}
$\sigma_{1}$ [km/h]& $\mu_{1}$ [km/h] & $\sigma_{2}$ [km/h] & $\mu_{2}$ [km/h]\\
\hline	
139.1& 4.58 & 127.8 & 7.80
\end{tabular}
\end{ruledtabular}
\label{tb2}
\end{table}

\begin{figure}[h!]
\centering
\includegraphics[width=7cm]{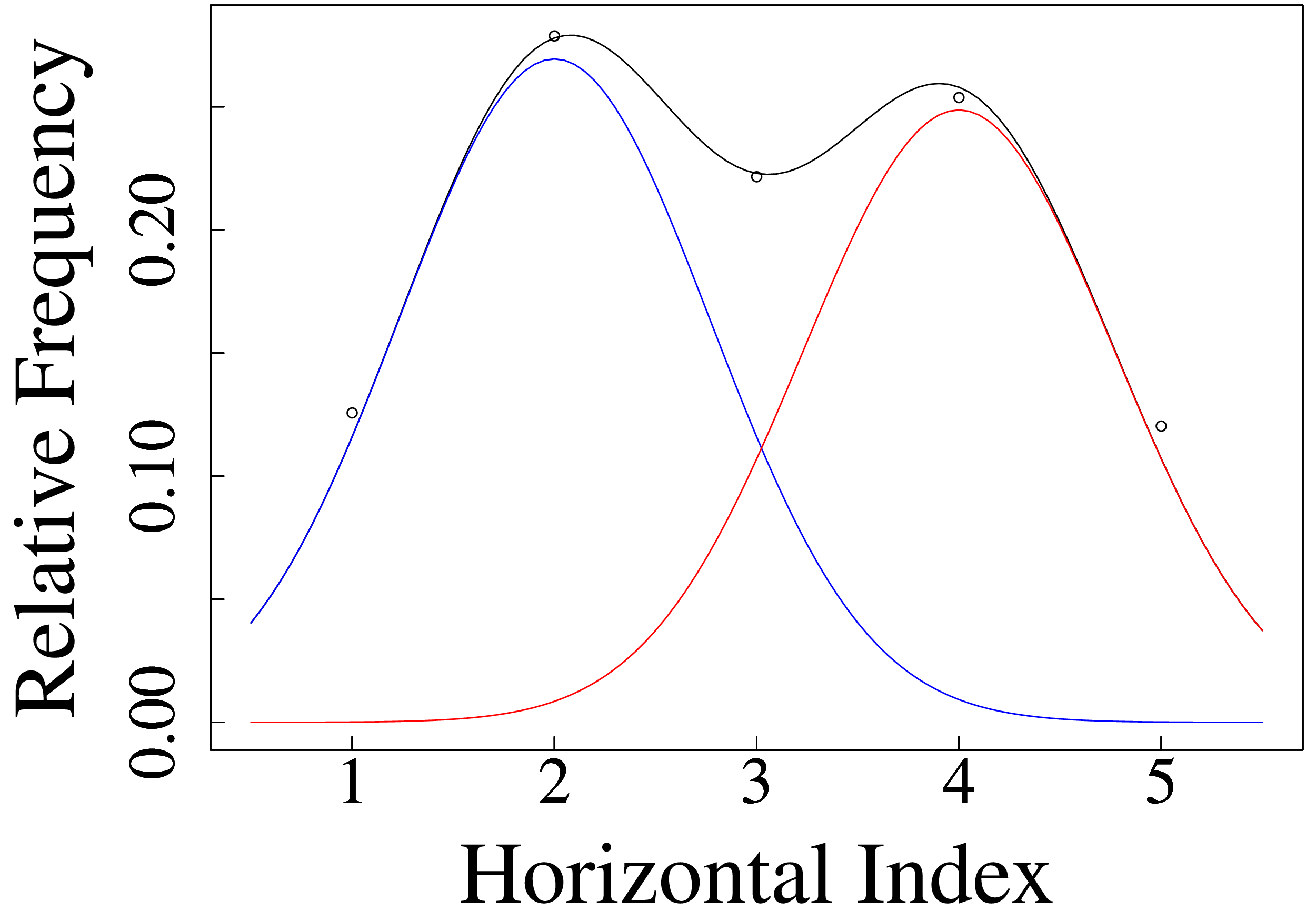}
\caption{Distribution of the horizontal position of pitched balls at the home base. 
Solid black curve shows Eq. (\ref{md2}) with $q=0.52$.}
\label{fig3}
\end{figure}

Next, we show the distributions of the horizontal position $y$ and 
the vertical position $z$ of the pitched balls in the pitching zone. 
Figure \ref{fig3} shows the distribution of the horizontal position of 
the pitched balls, where the horizontal axis indicates the positions 
of the grids in Fig. \ref{fig1}, which are numbered from left to right. 
The open circles indicate the calculated probabilities as a function 
of the horizontal position. To obtain the distribution function 
approximating these data, we fitted the mixture distribution 
of the normal distributions, which is expressed as 
\begin{eqnarray}
\psi(y) = q h_{1}(y) + (1-q) h_{2}(y) \hspace{3mm} (0 < q < 1), \label{md2}
\end{eqnarray} 
where $h_{1}(y)$ and $h_{2}(y)$ are the normal distributions with the same standard 
deviation: 0.77.  

The means of $h_{1}(y)$ and $h_{2}(y)$ are $y=2$ and $y=4$, respectively. 
In Fig. \ref{fig3}, the black solid curve shows Eq. (\ref{md2}) 
with the mixing parameter $q=0.52$, whereas the blue and red curves show the first 
and second terms, respectively, in the right-hand side of Eq. (\ref{md2}). The coefficient 
of determination of the fit is equal to 0.986, which indicate a good fit. 
This indicates that the pitchers tended to choose the right and left sides of 
the strike zone with a similar probability to prevent home runs.

\begin{figure}[h!]
\centering
\includegraphics[width=7cm]{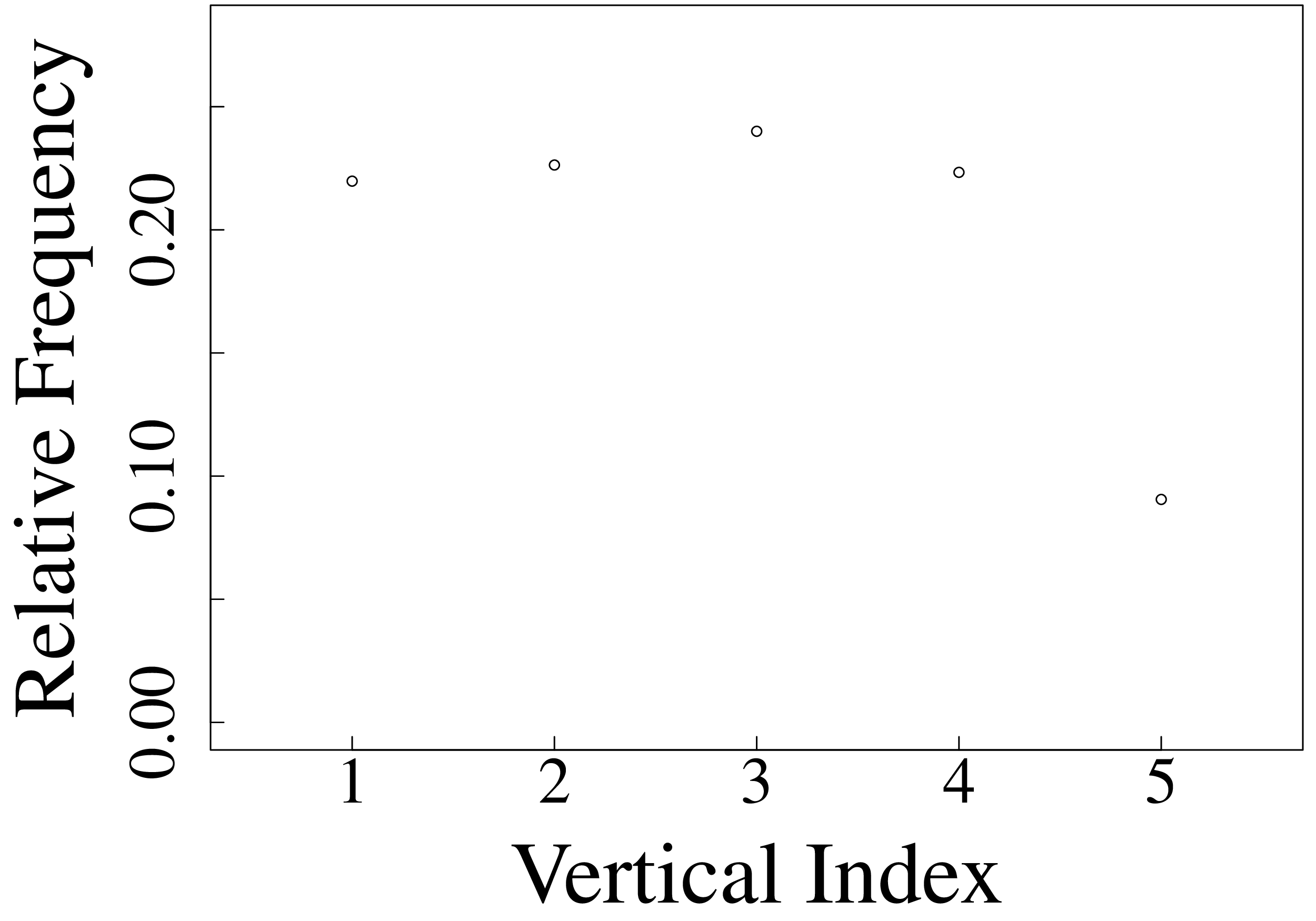}
\caption{Distribution of the vertical position of pitched balls at the home base. }
\label{fig4}
\end{figure}

On the other hand, Fig. \ref{fig4} shows the distribution of the vertical position 
for the pitched balls. 
The probability is almost constant, except at $z=5$, which is the highest position. 
This indicates that the pitchers tended to choose heights all over the strike zone 
with a similar probability. The frequency at $z=5$ is relatively small to avoid 
the risk of long ball hitting. On the other hand, the frequency at $z=1$ is almost same 
as those at $z=2, 3, 4$ inside the strike zone, which may be attributed to 
that the balls at $z=1$ are difficult to strike. 

\section{Simulation}

\begin{figure}[h!]
\centering
\includegraphics[width=12cm]{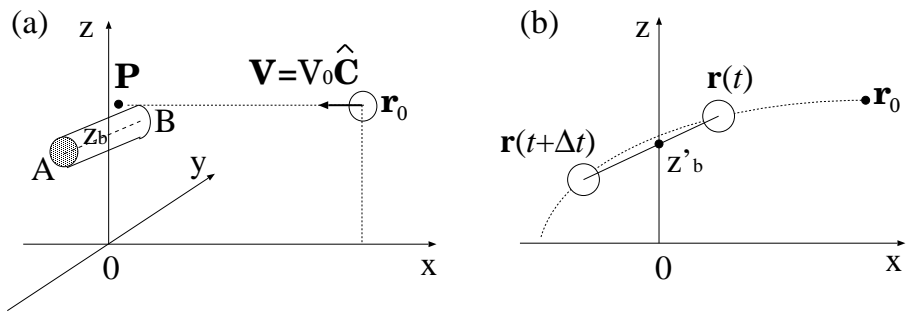}
\caption{Schematics showing (a) our simulation setup and (b) the definition of $z^{'}_{b}$.}
\label{fig5}
\end{figure}

We performed a batting simulation according to the results presented in the 
previous section. Figure \ref{fig5}(a) shows our simulation setup, 
in which a home plate was placed at the origin of a Cartesian coordinate system. 
The initial position of the center of mass of a baseball was set at 
${\bf r}_{0} = ($18.44 m, 0 m, 1.8 m$)$ according to the official baseball rules 
in Japan and the average height of Japanese professional baseball players. 
The baseball was pitched in the negative direction of the $x$ axis with 
an initial velocity of ${\bm V}_{0} =  V_{0} \hat { {\bm C} }$, 
where $\hat { {\bm C} }$ is the unit direction vector of the pitch, 
which will be defined later. Here, $V_{0}$ is randomly chosen according to 
the distribution function given by Eq. (\ref{md}) with the parameters shown 
in Table II. 

The pitch direction was defined as follows. A ball needs to be thrown from about shoulder height 
and be launched in an almost horizontal direction to cross the plate at the correct height\cite{cross}.  
Thus, we assume that a pitcher throws a ball toward a point ${\bf P} = ($0, $P_{y}$ [m], 1.8 m$)$ 
that is on the $y$-$z$ plane. A thrown ball travels a curved path to cross 
the plate at a height less than the initial height due to the gravitational force (Fig. \ref{fig5}(b)). 
$P_{y}$ is the random variable selected by the distribution function 
$\psi(P_{y})$ defined in Eq. (\ref{md2}). 
The unit vector of the pitch direction was defined as 
$\hat { {\bm C} } = {\bm C} / |{\bm C}|$, where ${\bm C} \equiv {\bf P} - {\bf r}_{0}$.  

\subsection{Modeling of Bat and Swing}

We consider the bat to be an uniform cylinder 1 m long and the diameter of the base 
to be $6.6 \times 10^{-2}$ m, in accordance with the official baseball rules. 
The bat is placed along the $y$ axis, and its center of mass is positioned 
at $($0, 0, $z_{b}$ [m]$)$, where $z_{b}$ is defined later. 
For later discussions, we respectively label bases of the bat as $A$ and $B$ so that 
the $y$ component of the center of $B$ is larger than that of $A$.

Let us assume that a thrown ball crosses the  $y$-$z$ plane between the time $t$ 
and $t+\Delta t$, where $\Delta t= 0.05$[s] is the time step of our simulation 
(Fig. \ref{fig5}(b)). First, we calculate $z_{b}^{'}$ using the $z$ component of 
the intersection between the  $y$-$z$ plane and the line segment connecting 
${\bf r}(t)$ and ${\bf r}(t+\Delta t)$, where ${\bf r}(t)$ is the 
position of the center of mass of a thrown ball at the time $t$. 
Next, we determine $z_{b}$ as follows: 
\begin{equation}
z_{b} = z_{b}^{'} + \sigma_{b},
\end{equation} 
where $\sigma_{b}$ represents the random numbers chosen from the normal distribution 
with a mean of $0$ and a standard deviation of $0.0366$ m, which is the diameter of 
an official baseball in Japan. 

\begin{figure}[h!]
\centering
\includegraphics[width=7cm]{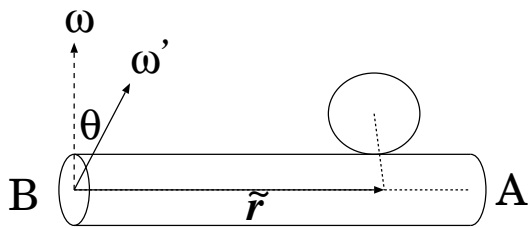}
\caption{A schematic of a collision between a ball and a bat.}
\label{sys}
\end{figure}
Figure \ref{sys} shows a schematic of a collision between a ball and a bat 
from the viewpoint of the home base. 
We assume the bat swings around an axis passing through the center of B. 
The bat swings with the angular velocity 
${\boldsymbol \omega}^{'} =(-\omega_{z} \sin \theta, 0, \omega_{z} \cos \theta)$ 
with $\omega_{z} =34$ rad/s, which is the typical value of bat swings. 
Here $\theta$ is the angle between the direction of ${\boldsymbol \omega}^{'}$ and 
${\boldsymbol \omega}=(0, 0, \omega_{z})$, the value of which is randomly chosen from the range 
$0^{\circ} \le \theta \le 10^{\circ}$.  
When a ball collides with the bat, we calculate the vector $\tilde{{\bf r}}$, which is defined 
by the vector from the center of  $B$ to the foot of the perpendicular line passing through 
the center of the ball to the central axis of the bat. 
The velocity of the bat on the collision is defined 
by ${\bf V}_{b} = {\boldsymbol \omega}^{'} \times \tilde{{\bf r}}$.

\subsection{Equation of Motion of Ball}

Basically, a pitched ball obeys the equation of motion 
\begin{eqnarray}\label{eqofmo}
m \frac{d^{2} {\bf r}}{dt^{2}} = - {\bf F}_{D} + {\bf F}_{L}  + m {\bf g},
\end{eqnarray}
where $m$ is the mass of the ball. 
The three terms on the right-hand side of Eq. (\ref{eqofmo}) represent the drag force, 
the magnus force, and the gravitational force, respectively.

The drag force ${\bf F}_{D}$ generated by the resistance from the air 
decreases the ball speed. ${\bf F}_{D}$ is expressed as 
\begin{eqnarray}
{\bf F}_{D} = \frac{1}{2} C_{D} \rho V^{2} A \hat{{\bf V}},
\end{eqnarray}
where $C_{D}$, $\rho$, $A$, and $\hat{{\bf V}}$ are the drag coefficient, 
the density of air, the cross section of a baseball, and the unit vector in the 
direction of the velocity, respectively. 
We use $C_{D} = 0.4$, corresponding to a baseball pitched at a high speed of 
approximately 40.2 to 44.7 m/s\cite{cross}. Our choice of the value will be 
discussed in later section. 
In addition, we use $\rho=1.29$ kg/m$^{3}$, 
which is the value at 0 C$^{\circ}$ and 1 atm.

On the other hand, the magnus force ${\bf F}_{L}$ generated by the backspin of 
the baseball enhances the flight of the ball 
after it is thrown and batted\cite{sawicki, nathan08}. ${\bf F}_{L}$ is expressed as 
\begin{eqnarray}
{\bf F}_{L} = \frac{1}{2} C_{L} \rho V^{2} A \hat{{\bf V}}_{n},
\end{eqnarray}
where $C_{L}$ and $\hat{{\bf V}}_{n} $ are the lift coefficient and 
the unit vector perpendicular to $\hat{{\bf V}}$, respectively. 
We used $C_{L}=0.2$, which is a typical value for a spinning ball\cite{cross}. 
The gravitational acceleration is indicated by the vector 
${\bf g}=(0, 0, -9.8$ m/s$^{2}$). 
We numerically solved Eq. (\ref{eqofmo}) using the velocity Verlet scheme\cite{frenkel}. 

\subsection{Condition for Collision}

A thrown ball changes its flight direction when it collides with a bat. 
We assume that the ball collides with the bat when all of the following conditions are 
fulfilled:
\begin{enumerate}
\item The distance between the center of mass of the ball and the central axis 
of the bat is less than the sum of the radius of the ball and that of the bat.
\item The $y$ component of the ball's position, $r_{y}$, satisfies $-0.5$ m $< r_{y} < 0.5$ m.
\end{enumerate}
When these conditions are fulfilled, the ball is reflected by the averaged repulsive force during 
$\Delta t$, 
\begin{equation}\label{collision}
\bar{\bf F} = -\frac{\mu (1 + \tilde{e}) {\tilde {\bf V}} \cdot {\bf n}}{\Delta t} {\bf n},
\end{equation}
where $\mu$, ${\bf n}$ and ${\tilde {\bf V}}$ are 
the reduced mass of the ball and the bat, 
the normal unit vector that is perpendicular to the tangential plane between a 
colliding ball and a bat, 
and the relative velocity of the ball to the bat, ${\tilde {\bf V}}={\bf V} - {\bf V}_{b}$, 
respectively. 

In Eq. (\ref{collision}), we use the coefficient of restitution between a ball and a bat
 $\tilde{e}$, which is defined by
 \begin{equation}
 \tilde{e} = e \left(1 + \frac{m}{M_{\mathrm{eff}}}\right) + \frac{m}{M_{\mathrm{eff}}},
 \end{equation}
where $M_{\mathrm{eff}} = I / \tilde{r}$ with the moment of inertia $I$ 
of the bat around the center of B\cite{cross, nathan03}. 
The derivation of Eq. (\ref{collision}) and the algorithm for the collision are summarized in 
Appendices A and B, respectively. 

When a batted ball falls on the ground, we calculate the distance $D$ 
between the point of landing and the home plate.

\section{Simulation Results}

\begin{figure}[h!]
\centering
\includegraphics[width=9cm]{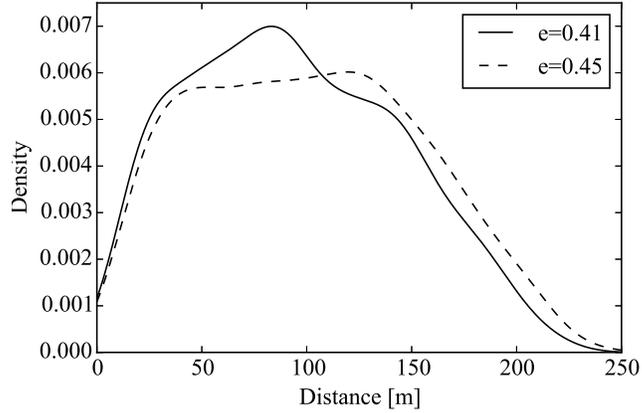}
\caption{Probability density of flying distance of batted balls. Solid and 
dashed curves show the results of $e=0.41$ and $e=0.45$, respectively.}
\label{fig6}
\end{figure}

Figure \ref{fig6} shows the probability densities of the flying distance $D$ of batted balls 
($e = 0.41$ and $e=0.45$), each of which was calculated from 1,000 samples. 
The highest peak position shifts from 75 m to 125 m with increasing coefficient of 
restitution of a ball.  The total frequencies in $D \ge 150$, which can be regarded as  
the number of home runs, increases with increasing coefficient of restitution indicating 
an increase in the home-run probability.
Thus, we investigate the relationship between the home-run probability and $e$ 
by changing the value of $e$ from $0.41$ to $0.45$. 

\begin{figure}[h]
\centering
\includegraphics[width=9cm]{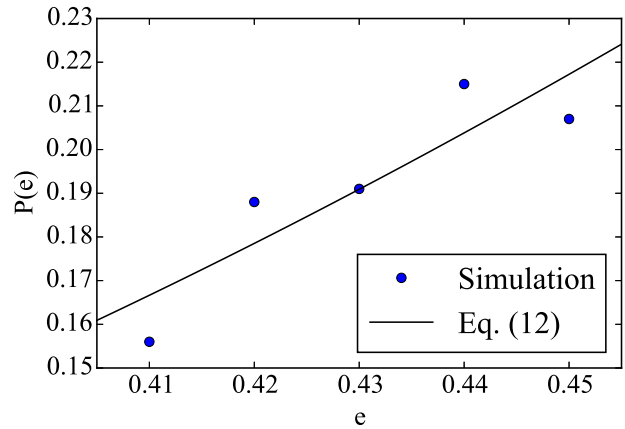}
\caption{Relationship between $e$ and home-run probability $P(e)$. Solid curve shows Eq. (\ref{pf_HR}) 
with $C_{1}=242$ m and $\sigma=52.2$ m. }
\label{fig7}
\end{figure}

Figure \ref{fig7} shows the relationship between the home-run probability and $e$. 
Here, each data point was calculated from 1,000 samples. 
We consider a sample with $D \ge 150$ m as a home run and define the home-run probability 
as the ratio of the number of home runs to 1,000. 
As shown in Fig. \ref{fig7}, the home-run probability 
increases with $e$, which is intuitively understood.

Here, we estimate the functional form of $P(e)$ using a simple theoretical argument. 
Let us suppose that a projectile is launched from the ground with a launch speed $v^{'}$ and 
a launch angle $\theta_{0}$ under the air friction.  
The range of the trajectory under the air friction can be estimated  
by a linear function of $v^{'}$ (Ref. 2), 
so that we roughly estimate the range of the trajectory as a 
linear function of the coefficient of restitution $e$ as $C_{1} e$ with the constant $C_{1}$. 
Here we assume that the speed of the thrown ball upon the collision with the bat 
is almost constant.

We assume that the probability density of the flying distance of a batted ball which will
land at approximately $D=150$ m can be approximated by the normal distribution as 
\begin{equation}\label{pdf_HR}
p(D) = \frac{1}{\sqrt{2 \pi \sigma^{2}}} \exp 
\left[ 
- \frac{(D-C_{1}e)^{2}}{2 \sigma^{2}}
\right].
\end{equation}
By integrating Eq. (\ref{pdf_HR}) from $D=150$ m to infinity, 
we obtain the probability function 
\begin{eqnarray}
P(e) &=& \int_{150}^{\infty} p(D) dD\\
&=& \frac{1}{2} 
\left[ 1 - {\rm erf} \left( \frac{150-C_{1} e}{\sqrt{2} \sigma}\right)
\right],\label{pf_HR}
\end{eqnarray}
where ${\rm erf}(x)$ is the error function. 
We assume both $C_{1}$ and $\sigma$ as 
fitting parameters. The solid curve in Fig. \ref{fig7} corresponds to 
Eq. (\ref{pf_HR}) drawn with 
$\sigma=59.8$ m and $C_{1}=271$ m, which indicates a good estimation.

\section{Discussion}

Figure \ref{fig3} shows the distribution of the horizontal index $y$ 
of pitched balls at the pitching zone, 
which can be approximated by the mixture distribution of the normal distributions. 
Because of the method of division of the pitching zone in the database that we used, 
the number of points was too scarce to identify a more detailed distribution. 
To obtain a more detailed distribution of the position of pitched balls, we can use more 
accurate data from PITCHf/x\cite{cross}. 
However, from the viewpoint of pitcher strategy, it may be desirable to aim 
at the edges of the strike zone; thus, we may find two peaks at the right 
and the left edges of the strike zone in the real distribution.

Figure \ref{fig6} shows the probability density distributions $p(D)$ of the distance $D$, 
each of which has some peaks.  
For comparison with real data, a Japanese group attempted to draw a histogram 
of the distance of batted balls 
using data from Ultimate Zone Rating (UZR)\cite{uzr} in 2009 and 2010. 
Their results show that the histogram has two peaks around 
$40$ and $100$ m, which is qualitatively close to our simulation results at $e=0.45$. 
However, their results show that the probability of finding a ball at the distance 
$40 \le D \le 80$ is very low,  although our simulation results show that 
the probability density  at $e=0.45$ is almost same in the same range. 
Notice that the group constructed the histogram using the positions at which 
the fielders caught or picked up the batted balls, which are sometimes different from 
the points of landing. 
Thus, their histogram can be considered as the mixture of the histograms of the position 
of pickup by the infielders and the outfielders, which may cause the discrepancy 
between their results and our results. 

In the UZR analysis, the probability of finding a batted ball between 91.44 and 121.92 m 
was 0.179. 
In our simulation, on the other hand, the home-run probability $P(e)$ was $\thicksim 0.17$ at $e=0.42$,  
which is the closest value to the averaged coefficient of restitution in 2010. 
The discrepancy between the two values is attributed to that the data in 2009 and 2010 were included in the UZR analysis. 

Finally, our model is a simplified model based on simple rigid-body collisions, which ignores 
the rotation of a batted ball. Our model includes the lift coefficient $C_{L}=0.2$ as a fixed value, 
although $C_{L}$ can vary depending on the rotation of the ball\cite{cross, sawicki}. 
For example, recent experiments have demonstrated that $C_{L}$ strongly depends on the 
spin parameter, $S \equiv R \omega/V$, where $\omega$ and $R$ are the angular velocity and 
the radius of a ball, respectively\cite{nathan08, kensrud_thesis, kensrud_cl}. 
On the other hand, $C_{D}$ also varies according to Reynolds number, 
although our model includes the drag coefficient $C_{D}=0.4$ 
as a fixed value. Especially, the drag force on a baseball is known to drop sharply 
at speeds typical for thrown and batted balls\cite{nathan08, kensrud_thesis, kensrud_cd, frohlich}. 
To improve our model, we will need to incorporate $C_{L}$ and $C_{D}$ dependant on the motion 
of thrown and batted balls. In addition, constructing a model based on elastic collisions in which 
the friction between the ball and bat is considered will yield more accurate results.

\section{Concluding Remarks}

In this paper, we analyzed real data for the speed and the course of pitched balls 
in professional baseball games in Japan. Our results show that the distribution 
of the ball speed can be approximated by the mixture distribution of two normal distributions. 
In addition, we found that the horizontal position of pitched balls in the pitching zone 
obeys the mixture distribution of two normal distributions, each of which has a peak at the edge 
of the strike zone.

We simulated collisions between baseballs and a bat, where the statistics of the pitching 
obey our analyzed results. We finally obtained the probability density distribution 
$p(D)$ of the distance $D$ between the home plate and the landing point of the batted balls 
to calculate the home-run probability as a function of the coefficient of restitution.  
By using a simple theoretical argument 
with the assumption that $p(D)$ around $D=150$ m can be approximated by the normal distribution, 
we quantified the home-run probability as a function of the coefficient of restitution $e$ 
of baseballs. 

As stated in the previous section, we will obtain more accurate results 
in future works by improving our model. Developing a methodology to calculate 
the home-run probability may yield useful information for designing baseballs.

\appendix   
\section{AVERAGED FORCE IN BINARY COLLISION OF RIGID BODIES}
In Ref.1, the author derived the rebound velocity of a ball in a head-on collision with a bat\cite{cross}. 
In this appendix, we derive the rebound velocity of a batted ball in a three dimensional oblique collision. 
Note that we ignore the rotation of a ball and a bat in our argument. 

\begin{figure}[h]
\centering
\includegraphics[width=6cm]{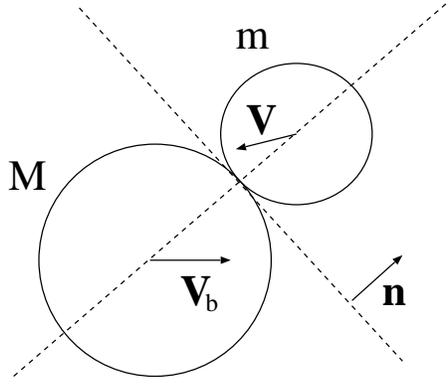}
\caption{
A schematic of a collision between a bat of mass $M$ and a ball of mass $m$. 
The velocities of the centers of mass of the ball and bat are denoted by ${\bf V}$ 
and ${\bf V}_{b}$, respectively. ${\bf n}$ is the normal unit vector perpendicular to 
the tangential plane.
}
\label{figa1}
\end{figure}

Figure \ref{figa1} shows a ball of mass $m$ colliding with a bat of mass $M$. Here we 
denote the colliding velocities of the ball and the bat as ${\bf V}$ and ${\bf V}_{b}$, respectively. 
Assuming that the ball experiences the averaged force ${\bar {\bf F}}$ from the bat during 
the duration $\Delta t$, we can describe the rebound velocities of the ball and the bat, 
${\bf V}^{'}$ and ${\bf V}^{'}_{b}$, as
\begin{eqnarray}
{\bf V}^{'} &=& {\bf V} + \frac{\bar{{\bf F}}}{m} \Delta t,\label{a1}\\
{\bf V}_{b}^{'} &=& {\bf V}_{b} - \frac{\bar{{\bf F}}}{M} \Delta t \label{a2}
\end{eqnarray}
from the definition of the averaged force and Newton's third law of motion. 
Note that the total momentum of the system is conserved after collision, 
$M {\bf V}_{b} + m {\bf V} = M {\bf V}_{b}^{'} + m {\bf V}^{'}$. 

By subtracting Eq. (\ref{a2}) from Eq. (\ref{a1}) and introducing the reduced mass $\mu $ 
defined by $1/\mu = 1/m + 1/M$, we obtain
\begin{eqnarray}
{\tilde {\bf V}}^{'}= {\tilde {\bf V}}  + \frac{1}{\mu} \bar{{\bf F}} \Delta t,
\end{eqnarray}
where ${\tilde {\bf V}}$ and ${\tilde {\bf V}}^{'}$ are the relative velocities 
of the ball to the velocity of bat before and after collision, respectively. 
By introducing the normal unit vector  ${\bf n}$ perpendicular to the tangential plane of the bat and ball, 
the scalar projection of ${\tilde {\bf V}}^{'}$ onto ${\bf n}$ is calculated as
\begin{eqnarray}
{\tilde {\bf V}}^{'} \cdot {\bf n} = {\tilde {\bf V}} \cdot {\bf n} + \frac{1}{\mu} \bar{F} \Delta t,\label{a4}
\end{eqnarray}
where we used $\bar{\bf F} = \bar{F} {\bf n}$.

Using Eq. (\ref{a4}) in the definition of the coefficient of restitution $\tilde{e}$ between ball and bat,  
\begin{equation}
| {\tilde {\bf V}}^{'} \cdot {\bf n}| = \tilde{e} | {\tilde {\bf V}} \cdot {\bf n}|,
\end{equation}
we obtain the equation
\begin{equation}
\left[ \beta + {\tilde {\bf V}} \cdot {\bf n} (1-\tilde{e})\right] 
\left[ \beta + {\tilde {\bf V}} \cdot {\bf n} (1+\tilde{e})\right] = 0,\label{qd}
\end{equation}
 where $\beta = \bar{F} \Delta t /\mu$. The two solutions of the quadratic equation Eq. (\ref{qd}) are 
 respectively written as
 \begin{eqnarray}
 \beta_{1} &=& - {\tilde {\bf V}} \cdot {\bf n} (1-\tilde{e}),  \\
 \beta_{2} &=& - {\tilde {\bf V}} \cdot {\bf n} (1+\tilde{e}). 
 \end{eqnarray}  
 In the two solutions, only $\beta_{2}$ corresponds to the averaged force in the collision 
 because $\beta_{1}$ becomes $0$ ($\bar{F} = 0$) when $e=1$, 
 which will cause the penetration of the ball into the bat.  

Thus, the averaged force is calculated as
\begin{eqnarray}
\bar{\bf F} = \bar{F} {\bf n} &=& \frac{\mu \beta_{2}}{\Delta t} {\bf n}\\
&=& -\frac{\mu (1 + \tilde{e}) {\tilde {\bf V}} \cdot {\bf n}}{\Delta t} {\bf n}. \label{a10}
\end{eqnarray} 

\section{ALGORITHM FOR COLLISION BETWEEN BALL AND BAT}

\begin{figure}[h]
\centering
\includegraphics[width=6cm]{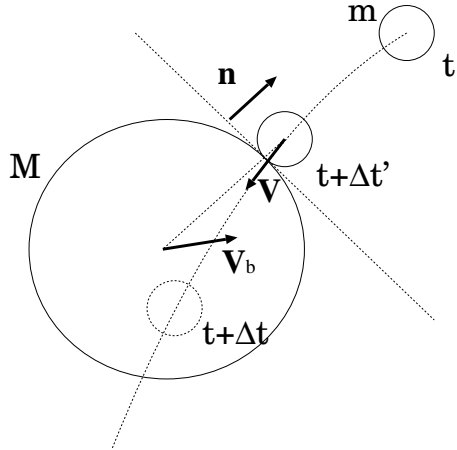}
\caption{
A schematic of a collision between a bat (large circle) and a ball (small circle). 
The ball at time $t$ will penetrate into the bat at the time $t + \Delta t$, which is 
unrealistic. To obtain the normal force acting on the ball, we need to calculate the time 
$t + \Delta t^{'}$ when the ball touches the bat. 
}
\label{figa2}
\end{figure}

In appendix A, we explained the way to calculate the averaged force acting on the ball 
during collision. However, it is difficult to obtain the unit normal vector ${\bf n}$ in naive 
calculation because the ball can penetrate into the bat during the simulation time step, which is 
due to the nature of the finite difference approximation of derivatives. 
In this appendix, we explain the algorithm of the collision between a bat and a ball to avoid penetration 
used in our simulation.

Figure \ref{figa2} shows a schematic of a collision between a ball and a bat, 
where a ball at the time $t$ is going to collide with the bat. When the first condition 
of collision (see section III.C) is fulfilled, the ball has penetrated into the bat at the time $t + \Delta t$. 
Thus, we put the ball back to the previous position at the time $t$ to determine the remaining time 
$\Delta t^{'}$ before collision. 
From the condition that the distance between the center of the bat and the center of the ball at 
the time $t + \Delta t^{'}$ equals to the sum of the radius of the bat and that of the ball, we obtain 
a quadratic equation for $\Delta t^{'}$. 
Here we assume that the ball travels linearly with a constant velocity during $\Delta t^{'}$ 
to simplify the calculation. 
Among the two solutions for the quadratic equation, we choose 
the positive one for $\Delta t^{'}$ which has a physical meaning. 

From the position of the ball at the time $t + \Delta t^{'}$, 
we can calculate all the variables such as ${\bf n}$, $\tilde{\bf r}$, ${\bf V}_{b}$ 
and $\tilde{e}$ in Eq. (\ref{collision}). The ball reflects from the bat according to 
the following algorithm. First, we put the ball back to the position at the time $t$. 
Next, we apply the half of the averaged force to the ball during $2 \Delta t$ so that 
the impulse from the bat becomes constant. 
With this two-step time evolution, the ball bounces from the bat without penetration.  

\begin{acknowledgments}
We gratefully acknowledge Data Stadium for permitting our use of their data. 
\end{acknowledgments}


\begin{thebibliography}{99}

\bibitem{cross} R. Cross, \textit{Physics of Baseball \& Softball}, 1st edition (Springer, New York, 2011).

\bibitem{adair} R. K. Adair, \textit{The Physics of Baseball}, 3rd edition (Harper Perennial, New York, 2002).

\bibitem{stronge} W. J. Stronge, \textit{Impact Mechanics}, 1st edition (Cambridge Univ. Press, Cambridge, 2004).

\bibitem{johnson} K. L. Johnson, \textit{Contact Mechanics}, 1st edition (Cambridge Univ. Press, Cambridge, 1985).

\bibitem{goldsmith} W. Goldsmith, \textit{Impact -The Theory and Physical Behavior of Colliding Solids-}, 
1st edition (Edward Arnold (Publishers) LTD., London, 1960).


\bibitem{louge} M. Y. Louge and M. E. Adams, Phys. Rev. E {\bf 65}, 021303-1--021303-6 (2002).


\bibitem{kuninaka_prl} H. Kuninaka and H. Hayakawa, ``Anomalous Behavior of Coefficient of Normal Restitution 
in Oblique Impact", Phys. Rev. Lett. {\bf 93}, 154301-1--154301-4 (2004).

\bibitem{drane} P. J. Drane and J. A. Sherwood, ``Characterization of the effect of temperature on baseball COR 
performance", 5th International Conference on the Engineering of Sport (UC Davis, CA), \textbf{2}, 59-65 (2004).

\bibitem{allen} T. Allen, et al., ``Effect of temperature on golf dynamics'', 
Procedia Engineering, \textbf{34}, 634-639 (2012).

\bibitem{kagan} D. Kagan and D. Atkinson, ``The Coefficient of Restitution of Baseballs as a 
Function of Relative Humidity'',  
Phys. Teach. \textbf{42}, 330--354 (2004).  

\bibitem{kagan2}
D. T. Kagan, ``The effects of coefficient of restitution variations on long fly balls", 
Am. J. Phys., \textbf{58}, 151--154 (1990).

\bibitem{npb}
Nippon Professional Baseball Organization, \url{<http://www.npb.or.jp>}.

\bibitem{brc}
Baseball-Reference.com, \url{<http://baseball-reference.com>}.

\bibitem{sawicki} G. S. Sawicki, M. Hubbard, and W. J. Stronge,  
``How to hit home runs: Optimum baseball bat swing parameters for maximum range trajectories", 
Am. J. Phys., \textbf{71}, 1152--1162 (2003).

\bibitem{nathan11} A. M. Nathan, L. V. Smith, W. M. Faber, and D. A. Russell, 
``Corked bats, juiced balls, and humidors: The physics of cheating in baseball", 
Am. J. Phys., \textbf{79}, 575--580 (2011).

\bibitem{sportsnavi} 
Sportsnavi, \url{<http://baseball.yahoo.co.jp/npb/>}.

\bibitem{bat}
M. Kasahara {\it et al}.,
`` Factors affecting on bat swing speed of university baseball players", 
NSCA Jpn J., \textbf{19}(6), 14--18 (2012). 

\bibitem{nathan08}
A. M. Nathan, ``The effect of spin on the flight of a baseball", 
Am. J. Phys., \textbf{76}, 119--124 (2008).

\bibitem{frenkel} D. Frenkel and B. Smit, \textit{Understanding Molecular Simulation}, 2nd edition (Academic Press, San Diego, 2002).

\bibitem{nathan03}
A. M. Nathan, ``Characterizing the performance of baseball bats", Am. J. Phys., \textbf{71}, 134--143 (2003).

\bibitem{uzr}
Baseball Lab. archives (in Japanese), 
\url{https://web.archive.org/web/20131230140449/http://archive.baseball-lab.jp/column_detail/&blog_id=17&id=155}. 


\bibitem{kensrud_thesis}
J. R. Kensrud, ``Determining Aerodynamic Properties of Sports Balls in Situ", MS Thesis, Washington State Univ., 2010.

\bibitem{kensrud_cl}
J. R. Kensrud and L. V. Smith, ``In situ lift mesuarement of sports balls", Proc. Eng., \textbf{13}, 278--283 (2011). 

\bibitem{kensrud_cd}
J. R. Kensrud and L. V. Smith, ``In situ drag mesuarement of sports balls", Proc. Eng., \textbf{2}, 2437--2442 (2010). 

\bibitem{frohlich}
C. Frohlich, ``Aerodynamic Drag Crisis and Its Possible Effect on the Flight of Baseballs", Am. J. Phys., 
\textbf{52}, 325--334 (1984). 

\end{thebibliography}
\end{document}